\documentclass[aps,prd,a4paper,twocolumn,floats,showpacs,superscriptaddress]{revtex4} %

\usepackage{amsmath,amssymb}
\usepackage{placeins,float}
\usepackage{color,epsfig,rotate,graphicx}
\usepackage{blindtext}
\usepackage{hyperref} 
 
\def\nn{\nonumber\\}
\def\beq{\begin{equation}}
\def\eeq{\end{equation}}
\def\bea{\begin{eqnarray}}
\def\eea{\end{eqnarray}}
\def\bal{\begin{align}}
\def\eal{\end{align}}

\def\kms{~{\rm km s^{-1}}}
\def\mpc{{\rm Mpc^{-1}}}

\def\ie{{\em i.e.}}
\def\eg{{\em e.g.}}

\def\fig{{\rm Fig.}}
\def\bwt{\begin{widetext}}
\def\ewt{\end{widetext}}

\def\LCDM{$\Lambda${\rm CDM}}
\def\nin{\noindent} 
\def\dsum{\displaystyle\sum}

\graphicspath{{GR_Effects_in_IDE_Figs/}{../GR_Effects_in_IDE_Figs/}}

\begin{document}

\title{Probing  the imprint of interacting dark energy on very large scales}
\author{Didam G. A. Duniya}
\affiliation{Physics Department, University of the Western Cape, Cape Town 7535, South Africa}

\author{Daniele Bertacca}
\affiliation{Physics Department, University of the Western Cape, Cape Town 7535, South Africa}
\affiliation{Argelander-Institut f\"ur Astronomie, Auf dem H\"ugel 71, D-53121 Bonn, Germany}

\author{Roy Maartens}
\affiliation{Physics Department, University of the Western Cape, Cape Town 7535, South Africa}
\affiliation{Institute of Cosmology \& Gravitation, University of Portsmouth, Portsmouth PO1 3FX, UK}

\date{\today} 

\begin{abstract}\nin
The observed galaxy power spectrum acquires relativistic corrections from lightcone effects, and these corrections grow on very large scales. Future galaxy surveys in optical, infrared and radio bands will probe increasingly large wavelength modes and reach higher redshifts. In order to exploit the new data on large scales, an accurate analysis requires inclusion of the relativistic effects. This is especially the case for primordial non-Gaussianity and for extending tests of dark energy models to horizon scales. Here we investigate the latter, focusing on models where the dark energy interacts non-gravitationally with dark matter. Interaction in the dark sector can also lead to large-scale deviations in the power spectrum. If the relativistic effects are ignored, the imprint of interacting dark energy will be incorrectly identified and thus lead to a bias in constraints on interacting dark energy on very large scales. 
\end{abstract} 

\maketitle

\section{Introduction}\label{sec:intro}
There is no observational evidence yet  for the existence of non-gravitational interaction between dark energy (DE) and dark matter (DM), nor is there guidance from fundamental theory as to the possible forms of such an interaction. However, there is also no consistent explanation for standard (non-interacting) DE from fundamental theory. And interacting dark energy (IDE) is not ruled out by current observations.  (For a range of work on IDE, see  \cite{Chimento:2003iea}--\cite{Pace:2014tya}.)

Various observations have been used to place constraints  on IDE models. In nearly all cases, the IDE models are compatible with current observations, provided that the interaction is not too strong. A few of the models result in non-adiabatic large-scale instabilities (see, \eg\ \cite{Valiviita:2008iv, Honorez:2009xt}), and hence may be ruled out. However, several of these models can be `fixed' by adjusting the DE equation of state (see, \eg~\cite{Valiviita:2009nu,Majerotto:2009np,CalderaCabral:2009ja}).

As far as we are aware, there has not up to now been an analysis of the IDE imprints on structure formation that takes into account the general relativistic (GR) effects on the galaxy power spectrum. These GR effects arise from observing galaxies on the past lightcone, and they become significant on very large scales \cite{Yoo:2010jd}--\cite{Raccanelli:2013gja}. Future galaxy surveys will cover wide sky areas and reach high redshifts, and thus begin to probe scales beyond the the equality scale, and approaching the Hubble scale (at higher redshift $z$). In order to analyse observations on these scales, we need to use the correct theoretical model -- \ie, including the GR effects which deviate from the Newtonian approximation that is accurate on smaller scales. 

Future high-volume galaxy surveys will allow us to: (i) extend tests of DE and modified gravity models -- and general relativity itself -- to horizon scales; (ii) measure the primordial non-Gaussian signal in the galaxy power spectrum at higher precision levels than the CMB. GR effects have been discussed in the case of (i) by \cite{Hall:2012wd,Lombriser:2013aj,Duniya:2013eta} and in the case of (ii) by \cite{Yoo:2012se,Camera:2014bwa,Camera:2014sba,Bruni:2011ta,Jeong:2011as,Maartens:2012rh}. Recently, \cite{Hashim:2014rda} investigated the  degeneracy between the large-scale imprint of primordial non-Gaussianity and the imprint of some IDE models, but without considering the GR corrections to the power spectrum.

Here we extend the work on DE models by investigating the GR effects on the galaxy power spectrum in IDE, assuming primordial Gaussianity. We normalise each IDE model to give the same matter  density parameter $\Omega_{m0}$ and Hubble constant $H_0$ as its corresponding standard non-interacting model. Then scale-dependent deviations from standard behaviour are clearly isolated on large scales at $z=0$. For $z>0$ however, the power spectra do not match on small scales. We show these behaviours in subsection~\ref{subsec:GReffects}.

The rest of the paper is organised as follows. We describe the background IDE models in Section \ref{sec:IDEbackground} and the perturbed IDE models in Section \ref{sec:IDEperturbation}. We investigate and discuss the very large scale behaviour of the power spectrum, including GR corrections,  in Section \ref{sec:Pofk}. Then in Section \ref{sec:concl}, we conclude.

\section{Background Universe with IDE}\label{sec:IDEbackground}

In the background universe dominated by (cold) DM ($A=m$) and DE ($A=x$), the interaction is defined via energy density transfer rates $\bar{Q}_A$, given by
\bea\label{bkgnd:IDE1}
\bar{\rho}_A' + 3{\cal H} (1+w_A) \bar{\rho}_A = a\bar{Q}_A,
\eea
where an overbar denotes background, a prime denotes derivative with respect to conformal time $\eta$, and ${\cal H} = a'/a$ is the conformal Hubble rate. The equation of state parameters  are $w_m=\bar p_m/\bar\rho_m=0$ and $w_x = \bar{p}_x / \bar{\rho}_x$. The conservation of the total energy-momentum tensor, $\nabla_\mu\sum_A \bar{T}^{\mu\nu}_A =0$, then implies
\beq\label{Q:consv} 
\bar{Q}_m = -\bar{Q}_x.
\eeq
The case $\bar{Q}_x > 0$ corresponds to energy density transfer from DM to DE, and vice versa for $\bar{Q}_x < 0$. We may define effective equation of state parameters 
\bea\label{eff:EoS}
w_{A,{\rm eff}} \;\equiv\; w_A -\dfrac{a\bar{Q}_A}{3{\cal H}\bar{\rho}_A},
\eea
where
\beq
\bar{\rho}_A' + 3{\cal H}(1+w_{A,{\rm eff}}) \bar{\rho}_A = 0,
\eeq
so that $w_{A,{\rm eff}}$ encode the deviations from the standard evolution of the dark sector energy densities. 

The background field equations are determined by the total energy-momentum tensor $\sum_A \bar{T}^{\mu\nu}_A$, and thus do not explicitly contain the interaction:
\bea
{\cal H}^2={8\pi Ga^2\over 3}(\bar\rho_m+\bar\rho_x),~~{\cal H}'=-{1\over2}\big(1+3w_x\Omega_x\big){\cal H}^2,
\eea
where $\Omega_A \equiv \bar{\rho}_A/\bar{\rho}$ and $\bar{\rho}$ is the total background energy density. (Note that $\bar{\rho}_A$ implicitly contain the interaction.) We neglect baryons for simplicity -- this does not affect the subject of our investigation.

We discuss our various choices of $\bar Q_A$ in Section \ref{sec:IDEperturbation}.

\section{Perturbed Universe with IDE}\label{sec:IDEperturbation}%

The perturbed Friedmann-Robertson-Walker (FRW) metric in the conformal Newtonian gauge, with vanishing anisotropic stress, is given by
\beq\label{metric}
ds^2 = a^2 \left[-(1+2\Phi) d{\eta}^2 + (1-2\Phi) d{\bf x}^2 \right],
\eeq
where $\Phi$ is the gravitational potential. 
The perturbed field equations do not contain the interaction explicitly when written in terms of Newtonian-gauge quantities, since they are determined by the total perturbed energy-momentum tensor $\sum_A\delta{T}^{\mu\nu}_A$:
\bea\label{Phi:Eqn}
\Phi' + {\cal H}\Phi &=& -\dfrac{3}{2} {\cal H}^2 \dsum_A{\Omega_A (1+w_A)V_A},\\
\label{spe}
\nabla^2 \Phi &=& \dfrac{3}{2} {\cal H}^2 \dsum_A{\Omega_A\big[\delta_A - 3{\cal H}(1+w_A)V_A \big]}.
\eea
Here $\delta_A = \delta{\rho}_A/ \bar{\rho}_A$ and the velocity potentials $V_A$ give the dark sector peculiar velocities, where the 4-velocities are
\bea\label{overDens:Vels}
u^\mu_A = a^{-1}\left(1 -\Phi,\, \partial^i V_A\right), \quad u^\mu = a^{-1}\left(1 -\Phi,\, \partial^i V\right).
\eea
The total $4$-velocity is  $u^\mu$ and  $V$ is the total velocity potential, given by
\beq\label{V:Eqn}
\Big(1+\dsum_A{\Omega_A w_A}\Big)V = \dsum_A{\Omega_A (1+w_A)V_A}.
\eeq

It is convenient to use the comoving overdensities 
\beq \label{Delta_A}
\Delta_A=\delta_A+{\bar\rho_A' \over \bar\rho_A}V_A=\delta_A - 3{\cal H} (1 + w_{A,\rm eff}) V_A.
\eeq
For non-interacting DE, the relativistic Poisson equation \eqref{spe}  then becomes $\nabla^2 \Phi = \frac{3}{2} {\cal H}^2 \sum_A\Omega_A\Delta_A$. 

Note that the right hand side of the Poisson equation \eqref{spe} remains the same for both interacting and non-interacting DE, so that $\Phi$ is always determined by $\sum_A\Omega_A[\delta_A - 3{\cal H}(1+w_A)V_A]$.  In the case of interacting DE, we have from \eqref{Delta_A} that $\delta_A - 3{\cal H}(1+w_A)V_A = \Delta_A -  a\bar{Q}_A V_A / \bar{\rho}_A$, thus
\bea\label{Poisson:Eq}
\nabla^2 \Phi &=& \dfrac{3}{2} {\cal H}^2 \sum_A{\Omega_A\Big[\Delta_A - {a\bar{Q}_A \over \bar{\rho}_A} V_A \Big]}.
\eea

\subsection{Energy-momentum transfer $4$-vectors}\label{subsec:genIDE}%

Generally, an interacting system is defined by the energy-momentum balance equations, given by \cite{ Valiviita:2008iv, Koyama:2009gd, Clemson:2011an}
\beq\label{EM:balEqs}
\nabla_\nu{T}^{\mu\nu}_A = Q^\mu_A, \quad\quad \dsum_A{Q^\mu_A} = 0,
\eeq
where $Q^\mu_A$ are the energy-momentum transfer $4$-vectors 
\beq\label{gen:IDE}
Q^\mu_A = Q_A u^\mu + F^\mu_A, \;\; Q_A = \bar{Q}_A+\delta{Q_A}, \;\; u_\mu F^\mu_A = 0. 
\eeq
The energy density transfer rate $Q_A$ and the momentum density transfer rate $F^\mu_A$ are both relative to $u^\mu$. In first-order perturbations, we have
\beq\label{MRate}
F^\mu_A = a^{-1} \left(0,\; \partial^i f_A\right), 
\eeq
where $f_A$ is the momentum density transfer potential. Then  \eqref{overDens:Vels}, \eqref{V:Eqn} and \eqref{gen:IDE} imply that 
\bea
Q^A_0 &=& -a\left[\bar{Q}_A(1+\Phi) + \delta{Q}_A\right], \nonumber\\ Q^A_i &=& a\partial_i \left[\bar{Q}_A V + f_A\right].\label{Q0:Qi}
\eea

\subsection{General perturbed balance equations}\label{subsec:genEMeqs}%
By considering all species as perfect fluids, the perturbed energy-momentum tensor of species $A$ is given by
\bea\label{pertEMT}
\delta{T}^{\mu\nu}_A &=& \left(\delta\rho_A +\delta{p}_A\right)\bar{u}^\mu_A\bar{u}^\nu_A\nn
&& +\; \left(\bar{\rho}_A + \bar{p}_A\right)\left[\delta{u}^\mu_A\bar{u}^\nu_A + \bar{u}^\mu_A\delta{u}^\nu_A\right] \nn
&& +\; \delta{p}_A\bar{g}^{\mu\nu} + \bar{p}_A\delta{g}^{\mu\nu},
\eea
where $\delta{u}^\mu_A$ and $\delta{g}^{\mu\nu}$ are the perturbations in the $4$-velocity and the metric tensor, respectively. The pressure perturbation $\delta{p}_A$ is given by~\cite{Valiviita:2009nu}
\bea\label{delp}
\delta{p}_A = c^2_{sA}\delta{\rho}_A + 3{\cal H}(c^2_{aA} - c^2_{sA})\left(1+w_{A,\rm eff}\right)\bar{\rho}_A V_A,
\eea
where $c^2_{sA} = (\delta{p}_A / \delta{\rho}_A)_{\rm restframe}$ is the physical sound speed squared, and $c^2_{aA} = \bar{p}'_A / \bar{\rho}'_A$ is the square of the adiabatic sound speed.

Given \eqref{EM:balEqs}--\eqref{MRate}, \eqref{pertEMT} and \eqref{delp} we obtain the energy and the momentum balance equations:
\bea\label{dDel_Adt}
\Delta_A' &-& 3{\cal H}w_A\Delta_A + (1+w_A) \nabla^2 V_A \nn
&-& \dfrac{9}{2}{\cal H}^2 (1+w_A) \dsum_{B\neq A}{\Omega_B(1+w_B) \left[V_A - V_B\right]} \nn
&=& \dfrac{a}{\bar{\rho}_A} \left[\delta{Q}_A - 3{\cal H} f_A\right] \nn &+& \dfrac{a^2\bar{Q}_A}{(1+w_A)\bar{\rho}^2_A} \left[f_A +\bar{Q}_A(V-V_A) \right] \nn
&+& \dfrac{a\bar{Q}_A}{\bar{\rho}_A} \Big\lbrace 3{\cal H}(V_A-V) +\dfrac{\bar{Q}'_A}{\bar{Q}_A}V_A \nn &-&\Big[1 + {c^2_{sA} \over 1+w_A}\Big]\Delta_A \Big\rbrace ,\\ \label{dVel_Adt}
V_A' &+& {\cal H}V_A = -\Phi -\dfrac{c^2_{sA} \Delta_A}{1+w_A} \nn &+&\dfrac{a\bar{Q}_A\left(V - V_A\right) + af_A}{(1+w_A)\bar{\rho}_A}.
\eea
Here $c_{sm}=0$ and we take $c_{sx}=1$ (which is the value for quintessence). 

\subsection{Particular IDE models}\label{IDE:Cases}%

We  model DE as a fluid with constant equation of state $w_x$ and consider two interactions. We use $w$CDM to denote the non-interacting case and $iw$CDM the interacting case.

In these interaction models, we assume that  the transfer $4$-vectors $Q^\mu_A$ run parallel to the DE $4$-velocity:
\bea\label{trans:Case}
Q^\mu_x &= Q_x u^\mu_x =& -Q^\mu_m.
\eea 
This means that there is zero momentum transfer in the DE rest frame, which is the case for example in the models of~\cite{Amendola:2003wa,Amendola:2003eq,Xia:2009zzb,Baldi:2011wy,Potter:2011nv,Maccio:2003yk,Pace:2014tya}. From \eqref{trans:Case}, it follows that the momentum density transfer rates (relative to $u^\mu$) become
\beq\label{fm:fx}
f_x = \bar{Q}_x (V_x - V)=- f_m.
\eeq
For transfer $4$-vectors of the form \eqref{trans:Case}, the balance equations~\eqref{dDel_Adt} -- \eqref{dVel_Adt} lead to  
\bea\label{dDel_Adt:DM}
\Delta_m' &-& \dfrac{9}{2}{\cal H}^2 \Omega_x(1+w_x) \left[V_m - V_x\right] + \nabla^2 V_m \nn
&=& 3{\cal H}(V_m-V_x)\left[1  -\dfrac{a\bar{Q}_m}{3{\cal H}\bar{\rho}_m} \right] \dfrac{a\bar{Q}_m}{\bar{\rho}_m} \nn
&& +\;\dfrac{a}{\bar{\rho}_m} \left[\delta{Q}_m + \bar{Q}'_mV_m -\bar{Q}_m\Delta_m\right], \\\label{dVel_Adt:DM}
V_m' &+& {\cal H}V_m = -\Phi - \dfrac{a\bar{Q}_m}{\bar{\rho}_m} \left[V_m - V_x\right], 
\eea
for DM, and  
\bea\label{dDel_Adt:DE}
\Delta_x' &-& 3{\cal H}w_x\Delta_x - \dfrac{9}{2}{\cal H}^2 \Omega_m (1+w_x) \left[V_x - V_m\right] \nn
&=& -(1+w_x) \nabla^2 V_x + \dfrac{a}{\bar{\rho}_x} \left[\delta{Q}_x +\bar{Q}'_xV_x\right] \nn
&& -a \left[1 + {c^2_{sx} \over 1+w_x}\right] \dfrac{\bar{Q}_x}{\bar{\rho}_x}\Delta_x , \\ \label{dVel_Adt:DE}
V_x' &+& {\cal H}V_x  = -\Phi -\dfrac{c^2_{sx} }{1+w_x} \Delta_x ,
\eea
for DE. (We have left $c_{sx}$ unspecified for generality, but we set it to 1 for numerical solutions.)

To fully specify an IDE model, we need to define the $Q_A$, which we choose as follows.

\subsubsection*{Model~1: $Q_x \propto \rho_x$}\label{IDE:Mod1}%

We use a transfer rate \cite{Valiviita:2008iv, Clemson:2011an,Hashim:2014rda},
\beq\label{Mod1:Q}
Q_x = \Gamma \rho_x = \Gamma\bar{\rho}_x (1 +\delta_x)=-Q_m,
\eeq
where $\Gamma$ is a universal constant (i.e. it is fixed under perturbations). In the case $\Gamma<0$, this corresponds to decay of DE into DM. 

Then from  \eqref{overDens:Vels} and \eqref{Mod1:Q}, it follows that
\beq\label{Mod1:Qm4vec}
Q^x_\mu = a\Gamma \bar{\rho}_x \left[-(1+\delta_x + \Phi), \, \partial_i V_x\right] = -Q^m_\mu.
\eeq
By  \eqref{fm:fx},
\beq \label{Mod1:fmScalar}
f_x = \Gamma \bar{\rho}_x (V_x - V) = -f_m.
\eeq

\begin{figure*} 
\includegraphics[scale=0.45]{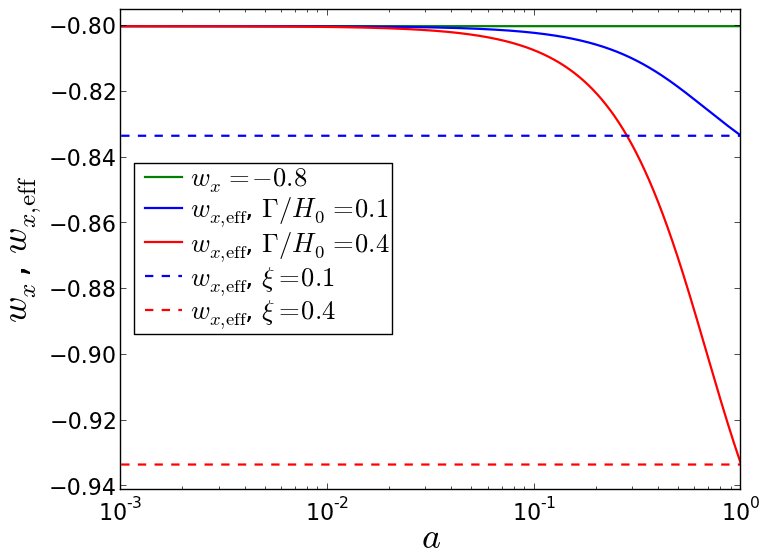} \includegraphics[scale=0.45]{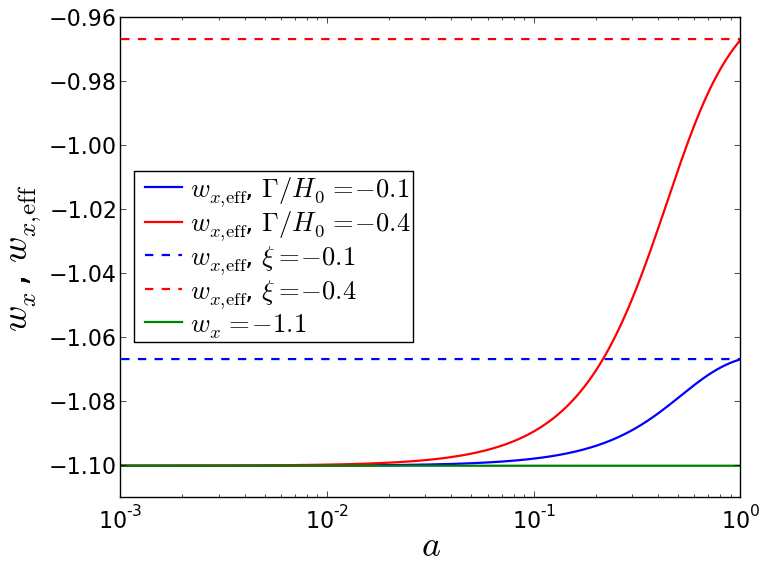}
\caption{Evolution of the IDE effective equation of state parameters $w_{x,\rm eff}$, for the $w$CDM equation of state parameters $w_x=-0.8$ ({\em left panel}) and $w_x=-1.1$ ({\em right panel}). Solid lines correspond to {\it Model~1}~\eqref{Mod1:Q} while dashed lines correspond to {\it Model~2}~\eqref{Mod2:Q}, and the $w_x$ line denotes $\Gamma = 0 = \xi$.}\label{fig:1}
\end{figure*} 

\subsubsection*{Model~2: $Q_x \propto \Theta\rho_x $}\label{IDE:Mod2}%

It is common in the literature to use energy density transfer rates of the form $Q_x=\xi a^{-1}{\cal H}\rho_x$, where $\xi=\,$const, \ie~to use a transfer rate proportional to the Hubble rate, rather than a constant rate $\Gamma$ as {\em Model 1} \eqref{Mod1:Q}. The motivation for this choice is that the background energy conservation equations are easily solved. However, the problem is that for the perturbed model, the Hubble rate ${\cal H}$ is typically not perturbed. 

We resolve this problem by using instead the self-consistent transfer rate
\beq\label{Mod2:Q}
Q_x = \dfrac{1}{3}\xi\rho_x \Theta,\quad \Theta = \nabla_\mu u^\mu.
\eeq
In the background, this reduces to the usual form, but in the perturbed universe we pick up additional perturbations of the expansion rate:
\beq\label{nabla:u}
\Theta = 3a^{-1}\left[{\cal H} - \left(\Phi' + {\cal H}\Phi\right) +\dfrac{1}{3}\nabla^2 V\right].
\eeq
This leads to 
\beq
 \label{Mod2:Q2}
Q_x = a^{-1}\xi{\cal H}\bar{\rho}_x \Big[1 +\delta_x -\Phi -\dfrac{1}{3{\cal H} }\left(3\Phi' - \nabla^2 V\right)\Big]=-Q_m.
\eeq
Then  \eqref{overDens:Vels}, \eqref{trans:Case}, \eqref{nabla:u} and \eqref{Mod2:Q2} imply that
\bea\label{Mod2:Qm4vec}
Q^x_\mu = \xi {\cal H} \bar{\rho}_x \left[ -1-\delta_x + \dfrac{1}{3{\cal H} }\left(3\Phi' - \nabla^2 V\right), \, \partial_i V_x\right].
\eea
By  \eqref{fm:fx}
\bea\label{Mod2:fmScalar}
f_x = a^{-1}\xi {\cal H} \bar{\rho}_x (V_x - V) = -f_m.
\eea~\\

For both models, the range of $w_x$ is restricted by stability requirements \cite{Clemson:2011an, Salvatelli:2013wra, Costa:2013sva}:
\beq\label{stab}
w_x > -1~~\mbox{for}~\xi,\Gamma > 0; ~~w_x < -1~~\mbox{for}~\xi,\Gamma < 0.
\eeq
These two cases correspond to different energy transfer directions, by \eqref{Mod1:Q} and \eqref{Mod2:Q}:
\beq\label{etd}
\mbox{DM $\to$ DE for}~\xi,\Gamma > 0; ~~~\mbox{DE $\to$ DM for}~\xi,\Gamma < 0.
\eeq

\subsection{Background evolution of the models}

\begin{figure*}
	\includegraphics[scale=0.45]{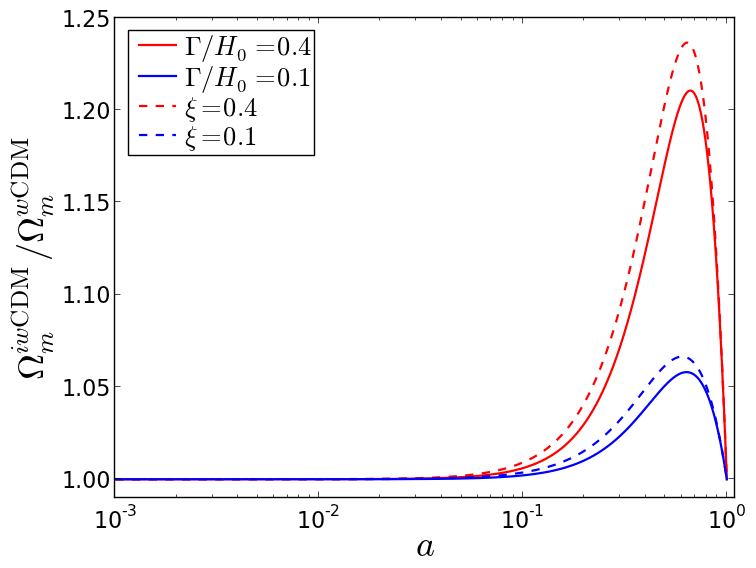} \includegraphics[scale=0.45]{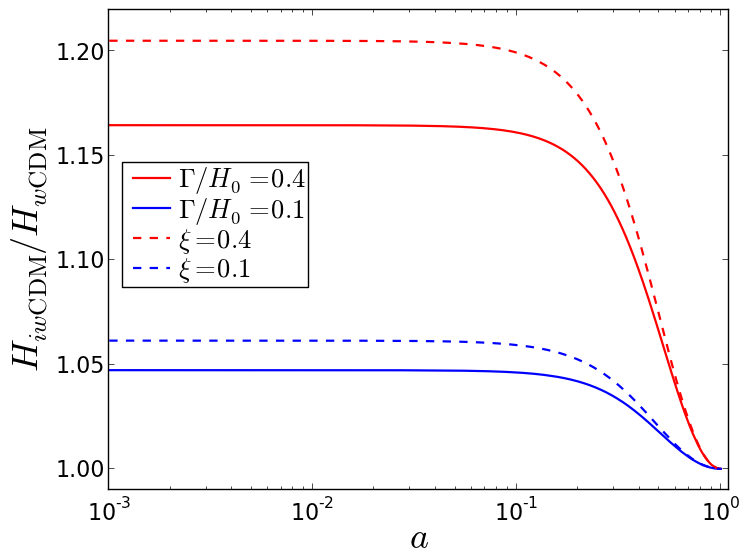}\\
	\includegraphics[scale=0.45]{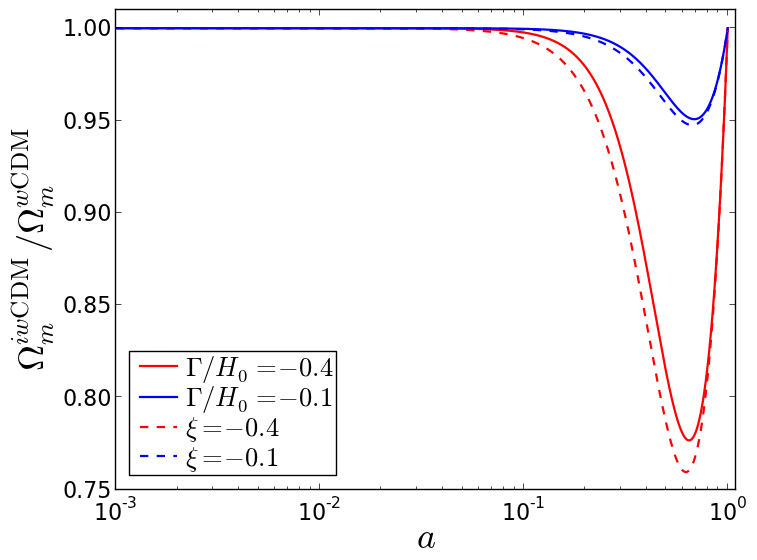} \includegraphics[scale=0.45]{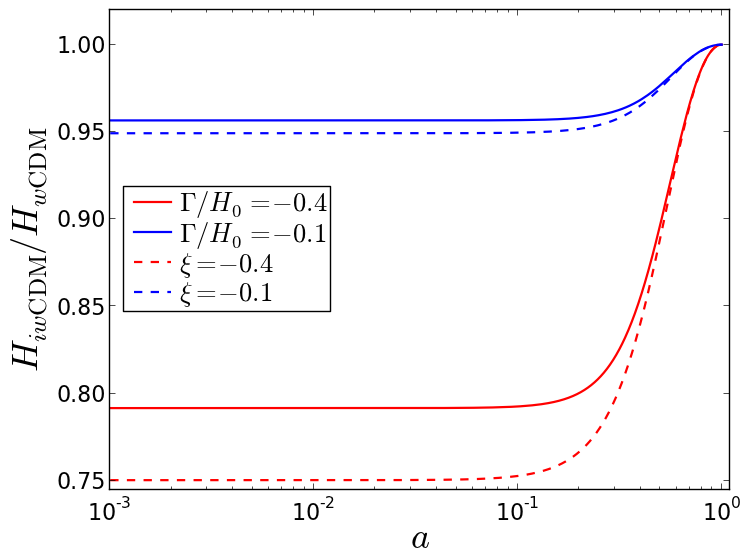}
\caption{Ratios of the matter density parameters ({\it left}) and Hubble rates ({\it right}) for $iw$CDM relative to those of $w$CDM: with $w_x=-0.8$ ({\em top panels}) and $w_x=-1.1$ ({\em bottom panels}). Line styles are as in Fig. \ref{fig:1}.}\label{fig:2}
\end{figure*}

We evolved the background equations from around decoupling, $a_d = 10^{-3}$, until today $a_0 =1$. The background initial conditions were chosen so that the matter density parameter $\Omega_{m0}$ and the Hubble constant $H_0$ are the same as in the non-interacting models. We used  $\Omega_{m0}=0.24$, $H_0=73\kms\mpc$. 

It is easy to understand the behaviour of the effective equation of state parameter $w_{x,\rm eff}$ in Fig.~\ref{fig:1}. For {\it Model~1} \eqref{Mod1:Q}, $w_{x,\rm eff} = w_x - a\Gamma/ (3 {\cal H})$, where $a {\cal H}^{-1}$ is a positive growing quantity. Hence $w_{x,\rm eff}$ gradually decreases for $\Gamma > 0$ ($w_x=-0.8$) and increases for $\Gamma < 0$  ($w_x=-1.1$), with $|\Gamma|$ determining the strength of the interaction (solid lines). However, for {\it Model~2} \eqref{Mod2:Q} we have $w_{x,\rm eff} = w_x - \xi/3$, which is a constant for constant $w_x$, shown in the dashed lines. 

Figure~\ref{fig:2} shows the evolution of the matter density parameters and Hubble rates, compared to the non-interacting case. When $|\bar{Q}_x|$ is small ($|\xi|,|\Gamma/H_0| \lesssim 0.1$),  $w_{x,\rm eff}$ is only slightly less (greater) than $w_x=-0.8$ ($w_x =-1.1$). This implies weaker (stronger) DE effects: the background matter density for $iw$CDM becomes enhanced (suppressed) relative to $w$CDM for $w_x =-0.8$ ($w_x = -1.1$). However, when the transfer rate is higher ($|\xi|, |\Gamma/H_0| \gtrsim 0.4$), $w_{x,\rm eff}$ is much smaller (bigger) than $w_x =-0.8$ ($w_x =-1.1$) and hence $iw$CDM has surplus (less) matter relative to $w$CDM for $w_x =-0.8$ ($w_x =-1.1$). 

Notice the distinct separation between the ratios of the Hubble rates. To understand this, we know that during matter domination, when $\Omega_m \approx 1$ ($\Omega_x \ll 1$), the ratio is constant. In this work we fixed the background initial conditions in $w$CDM and let those in $iw$CDM vary with each value of $\Gamma$ or $\xi$ so as to recover the same values of $\Omega_{m0}$ and $H_0$ as  in $w$CDM. As $\Gamma$ or $\xi$ vary, the initial conditions in $iw$CDM change, enough to amount to significantly differing initial amplitudes of the Hubble rates. The ratio does not evolve until DE domination,  converging at $a=1$ by our normalization. 
 
\section{The large-scale power spectrum}\label{sec:Pofk}

We probe the large-scale structure of the late-time Universe by relating the perturbations at late epochs to the primordial potential via growth functions and a transfer function (see subsection \ref{Growth:fns}), from which the matter power spectrum is computed.

\subsection{Linear growth functions}\label{Growth:fns}%

Here $\Phi_d$ is the gravitational potential at photon-baryon decoupling. It is related to the gravitational potential growth function $D_\Phi$ by \cite{Duniya:2013eta}:
\bea\label{eq:lps:3} 
\Phi(k,a) &=& {D_{\Phi}(k,a) \over a}\,\Phi_d(k),~~~ D_{\Phi}(k,a_d) = a_d,\\ \label{phid}
\Phi_d(k) &=& \frac{9}{10}\Phi_p(k) T(k),\\ \label{eq:lps:4}
\Phi_p(k) &=&   A\frac{\Omega_{m0}}{D_{\Phi 0}(k)}\, \left(\!\frac{k}{H_0}\!\right)^{(n-4)/2},
\eea
where $D_{\Phi 0}=D_\Phi(k,1)$, $\Phi_p$ is the primordial potential and $T(k)$ is the transfer function which accounts for perturbation evolution through radiation domination until radiation-matter transition. The constant $A=5\sqrt{2}\,\pi\delta_H/(3H^{3/2}_0)$ is the primordial amplitude of curvature perturbations, we adopt $\delta_H=5.6{\times}10^{-5}$ (see~\cite{Bonvin:2011bg}) which is the scalar amplitude at horizon crossing and $n=0.96$ is the scalar spectral index. 

The growth function $D_m$ of the comoving matter overdensity $\Delta_m$  describes the growth of (linear) matter perturbations after radiation-matter equality via \cite{Duniya:2013eta}
\beq\label{eq:lps:1}
\Delta_m(k,a) = -\frac{2}{3\Omega_{m0}}{k^2 \over H_0^2}D_m(k,a)\Phi_d(k).
\eeq
The matter velocity growth function $D_{V_m}$ is defined by
\beq\label{eq:lps:2}
V_m(k,a) = -\frac{2}{3\Omega_{m0}H_0^2}D_{V_m}(k,a)\Phi_d(k),
\eeq
where we have used \eqref{Phi:Eqn} and assumed matter domination and Einstein de Sitter regime (\ie~with $\Phi'=0$). Note that $D_m$ implicitly contains the interaction:
\beq\label{eq:Dm_eff}
D_m(k,a) ={D}^0_m(k,a) + \dfrac{a\bar{Q}_m(a)}{k^2\bar{\rho}_m(a)} D_{V_m}(k,a),
\eeq
where ${D}^0_m$ is the growth function in the standard non-interacting scenario. Equation~\eqref{eq:Dm_eff} is obtained by using \eqref{eq:lps:3} and~\eqref{eq:lps:1} in the Poisson equation~\eqref{Poisson:Eq} -- and assumed matter domination.

\begin{figure}[t]%
\includegraphics[scale=0.45]{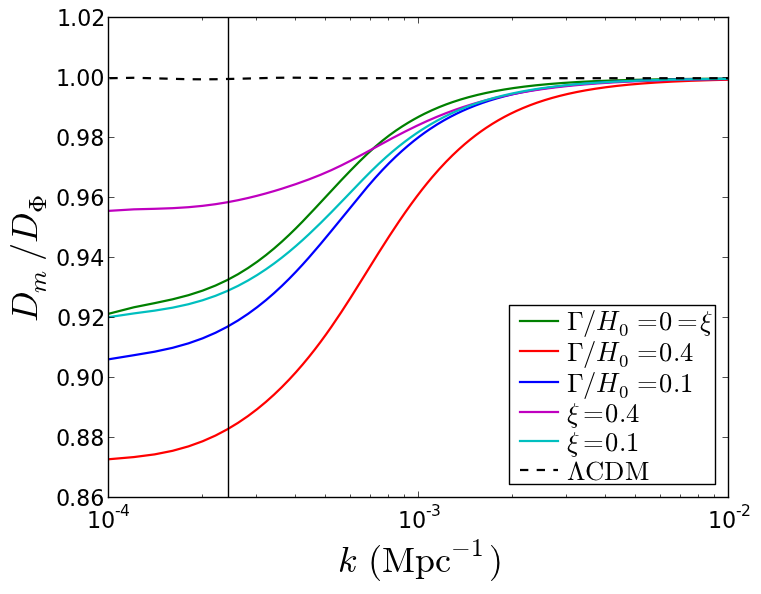}
\caption{The ratio of the matter overdensity and gravitational potential growth functions, at  $a_0=1$ (or $z=0$), with $w_x=-0.8$. Solid lines correspond to {\it Model~1}~\eqref{Mod1:Q} and dashed lines to {\it Model~2}~\eqref{Mod2:Q}. The  \LCDM\ case (dashed black line) and the Hubble horizon (solid black line) are also shown.}\label{fig:3}
\end{figure}
 
 \begin{figure*}
	\includegraphics[scale=0.45]{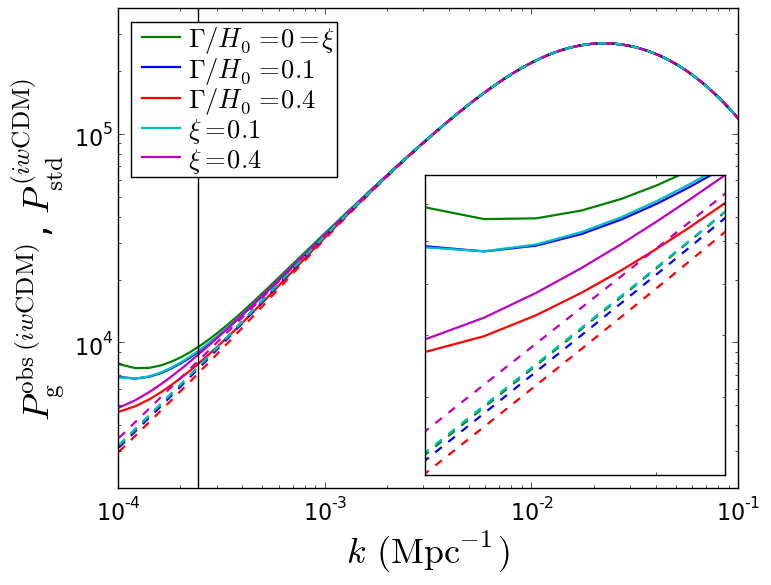} \includegraphics[scale=0.45]{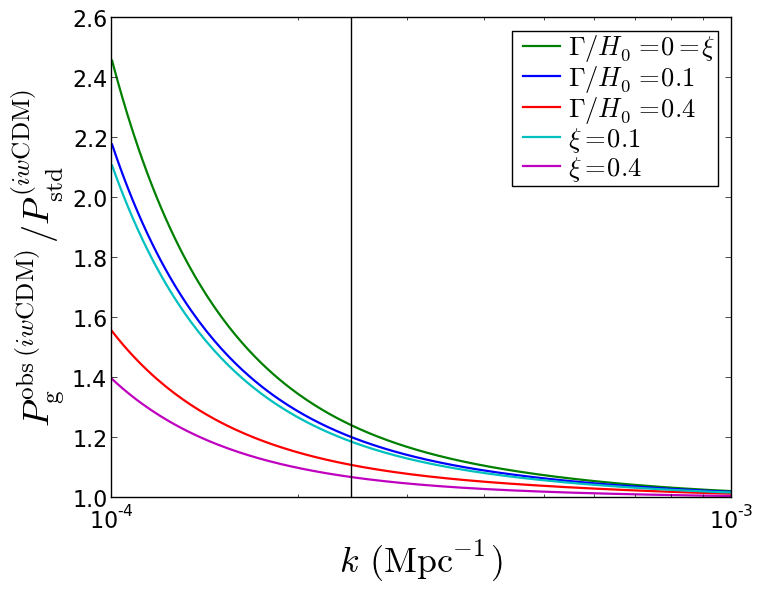} \\
	\includegraphics[scale=0.45]{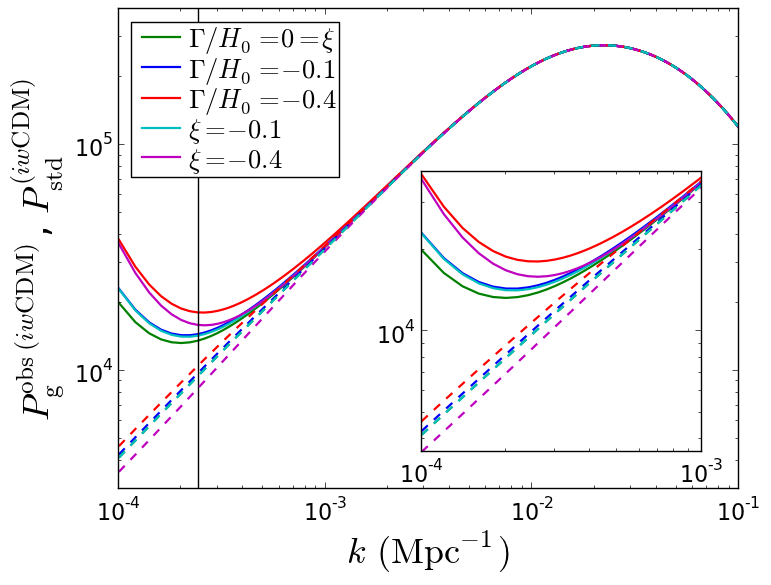} \includegraphics[scale=0.45]{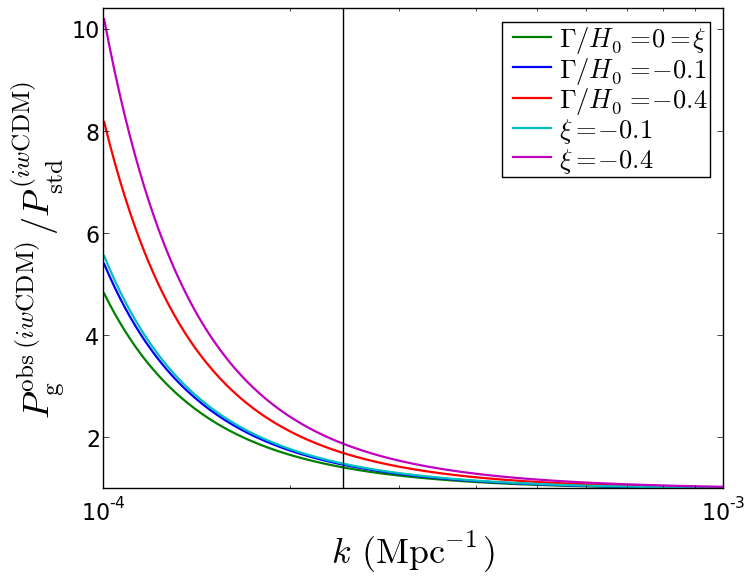}
\caption{Observed galaxy power spectrum $P^{\rm obs}_{\rm g}$ (solid lines) and the standard power spectrum $P^{\rm std}_{\rm g}$ (dashed lines) along the line of sight ($\mu=1$), at $z=0$: for $w_x=-0.8$ ({\it top}); and for $w_x=-1.1$ ({\it bottom}). The corresponding ratios of the power spectra are given on the right panels. }\label{fig:4}
\end{figure*} 

Note that $D_{\Phi}$ and $D_{V_m}$ in \eqref{eq:lps:3} and \eqref{eq:lps:2} retain the same definitions in both IDE and non-IDE scenarios -- \ie\ they do not explicitly contain the density transfer rate $\bar{Q}_{m}$.  In standard \LCDM, the growth functions become
\bea\label{dlcdm}
D_m(k,a)=D_\Phi(k,a) \equiv D(a),~D_{V_m}(k,a) = {dD(a) \over d\eta}.
\eea
However, for dynamical DE ($w_x\neq-1$), $D_m = D_\Phi$ only holds true on sub-Hubble scales $k\gg {\cal H}$.

The DE velocity growth function is related to that of matter by
\bea
D_{V_x}(k,a) &=& {\Omega_m(a) \over [1-\Omega_m(a)][1+w_x(a)]}\Big[{\partial  \over \partial\eta}D_\Phi(k,a) \nonumber\\ &&~ - D_{V_m}(k,a) \Big], \label{DV:DE}
\eea
where we have used \eqref{eq:lps:3} and \eqref{eq:lps:2} in \eqref{Phi:Eqn} and assuming Einstein de Sitter regime -- \ie~with $\Phi'=0$.

The DE overdensity growth function is 
\bea
D_x(k,a) &=& \dfrac{\Omega_m(a)}{1 - \Omega_m(a)} \Big[D_\Phi(k,a) - D_m(k,a) \nn 
&& - \dfrac{a\bar{Q}_m(a)}{k^2\bar{\rho}_m(a)}\Big\{D_{V_x}(k,a) - D_{V_m}(k,a)\Big\} \Big],\quad
\eea
where we used \eqref{eq:lps:3},~\eqref{eq:lps:2} and~\eqref{eq:lps:1} in the Poisson equation~\eqref{Poisson:Eq}, and the fact that $\bar{Q}_x/\bar{\rho}_x = -\Omega_m\bar{Q}_m/ (\Omega_x\bar{\rho}_m)$.

In \LCDM, by \eqref{dlcdm}, $D_x=0=D_{V_x}$. 

\subsection{Relativistic effects in the galaxy power spectrum}\label{subsec:GReffects}%

Using cosmological perturbation theory, we can describe the evolution and distribution of matter density perturbation in the Universe. However, in reality the matter perturbation distribution is not directly observable -- only objects such as galaxies, whose distribution traces that of the matter, are observable. The large-scale and scale-independent galaxy bias $b$ is usually defined by $\delta_{\rm g}=b\delta_m$. On sub-Hubble scales,  different gauge choices for $\delta_m$ agree, but on near- and super-Hubble scales, they disagree. The scale-independent bias needs to be defined physically, in the rest-frame of DM and galaxies (which coincide on large scales).
This leads to the following definition, valid on all linear scales (and assuming Gaussian primordial perturbations) \cite{Challinor:2011bk,Bruni:2011ta,Jeong:2011as}:
\beq\label{bias}
\Delta_{\rm g}(k,a) = b(a)\, \Delta_m(k,a).
\eeq

Furthermore, we do not observe in real (${\bf x}$) space, but in redshift space, leading to the Kaiser redshift-space distortion term:
\beq\label{kdo}
\Delta^{\rm obs}_{\rm g}(k,\mu,a) = \big[b(a) + f(k,a)\mu^2\big]\Delta_m(k,a),
\eeq
where 
\beq\label{grwth}
f \;=\; \dfrac{D_{V_m}}{{\cal H} D_m}
\eeq
which reduces to the growth rate of matter overdensity in non-interacting DE models, $\mu=-{\bf n}\cdot {\bf k}/k$ and ${\bf n}$ is the unit spatial vector in the direction of the photon geodesic $x^\mu(\lambda)$ from source to observer. Here $\lambda$ is the affine parameter along the photon geodesic, increasing from source to observer. (Note that $f$ in \eqref{grwth} is different from $f_m$ and $f_x$ in \eqref{fm:fx}.)

The Kaiser term is a flat-sky and sub-Hubble approximation to the full redshift-space distortion, which includes further velocity and Sachs-Wolfe type terms. In addition to the redshift distortion, there are other relativistic effects from observing galaxies on the past lightcone.  These include a contribution from weak lensing convergence, which can be significant on sub-Hubble scales at higher redshifts. In addition there are additional Sachs-Wolfe and Doppler terms, and integrated Sachs-Wolfe (ISW) and time-delay terms \cite{Yoo:2010jd,Yoo:2010ni,Bonvin:2011bg,Challinor:2011bk,Jeong:2011as,Bertacca:2012tp,Hu:2001yq}.

If we want an accurate analysis that includes near- and super-Hubble scales, we should use the galaxy overdensity that is observed on the lightcone, including all GR effects. This observed overdensity is automatically gauge-invariant. Here we will neglect the integrated terms and use a flat-sky approximation, generalizing the \LCDM\ form given in \cite{Jeong:2011as}:
\beq\label{Delta:Obs}
\Delta^{\rm obs}_{\rm g}(k,\mu,a) = \Delta^{\rm std}_{\rm g}(k,\mu,a) + \Delta^{\rm GR}_{\rm g}(k,\mu,a),
\eeq
where $\Delta^{\rm std}_{\rm g}$ is the standard term -- given by \eqref{kdo}. Note that in \eqref{Delta:Obs} we do not have a priori the time-delay, ISW and weak lensing integrated terms, and
\beq\label{Delta:GR}
\Delta^{\rm GR}_{\rm g} = \left[ {{\cal A}}{{\cal H}^2\over k^2} + i\mu{{\cal B}} {{\cal H}\over k}\right] \Delta_m.
\eeq
This form arises by using the field and conservation equations to relate velocity and potential to overdensity.
 (See \eg~\cite{Bertacca:2012tp,Yoo:2013tc,Duniya:2013eta} for the full GR expression including integrated terms.)  
The coefficients in \eqref{Delta:GR} are given in the \LCDM\ case by   \cite{Jeong:2011as}. We generalize their expressions for the case of IDE:
\bea 
{\cal A} &= & \left(3-b_e\right)f -\dfrac{3\Omega_{m0} H_0^2 }{2\, {\cal H}^2 D_m} \Big[4{\cal Q} - b_e - 1 + \frac{{\cal H}'}{{\cal H}^2} \nonumber\\ &&~
 + 2\frac{\left(1-{\cal Q}\right)}{r{\cal H}} + \dfrac{a^2}{D_\Phi} \dfrac{\partial }{\partial a}\Big({D_\Phi \over a}\Big)\Big] {D_\Phi \over a} ,\label{calA}\\ 
{\cal B} &= & \Big[ b_e -2{\cal Q}-\frac{\mathcal{H}'}{\mathcal{H}^2} \nonumber\\&&~
-2\frac{\left(1-{\cal Q}\right)}{r \mathcal{H}} + \dfrac{a\bar{Q}_m}{{\cal H}\bar{\rho}_m}\Big(1 - {D_{V_x} \over D_{V_m}}\Big) \Big] f. \label{calB}
\eea 
Here $r$ is the comoving radial distance at the observed galaxy, ${\cal Q}$ is the magnification bias,  and $b_e$ is the galaxy `evolution bias', giving the evolution of source counts (see \cite{Bonvin:2011bg,Challinor:2011bk,Jeong:2011as}). 

The interaction enters ${\cal B}$ through the last term in \eqref{calB}. This term arises from the perturbed Euler equation \eqref{dVel_Adt:DM}, which comes in via
\beq
\dfrac{1}{{\cal H}} \dfrac{d}{d\lambda}\big({\bf n}\cdot{\bf V}_m\big) = {1\over {\cal H}} {\bf n}\cdot{\bf V}_m'  + \dfrac{1}{{\cal H}} {n}^i\partial_i\big({\bf n}\cdot{\bf V}_m \big), 
\eeq
where the total derivative is taken along the photon geodesic, in the direction from source to observer. The $Q_m$ term  is absent in \LCDM\, \cite{Yoo:2010jd,Yoo:2010ni,Bonvin:2011bg,Challinor:2011bk,Jeong:2011as} and in non-interacting DE models~\cite{Duniya:2013eta}. It would also be absent in IDE models with $Q_A^\mu$ parallel to the matter 4-velocity $u_m^\mu$, since in these models the DM follows geodesics and the perturbed Euler equation is the same as for non-interacting DE.

\begin{figure*}
\begin{tabular}{cc}
\includegraphics[scale=0.45]{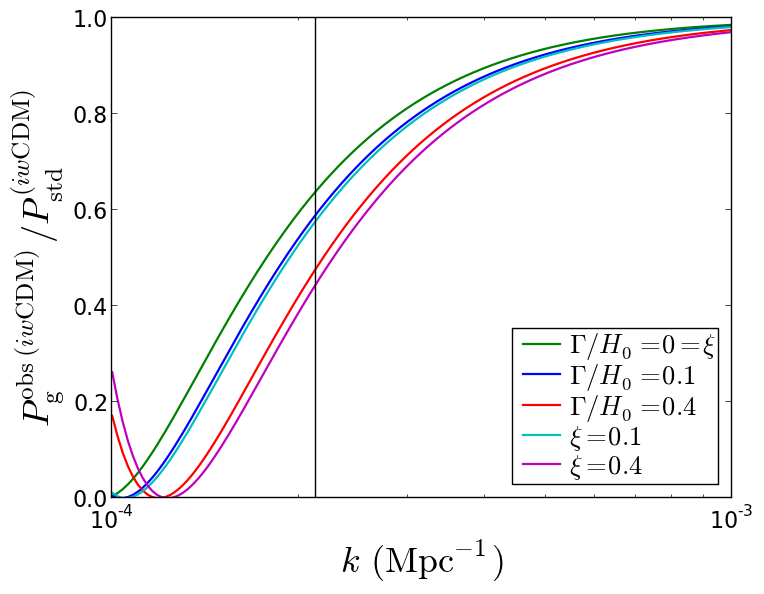} & 
\includegraphics[scale=0.45]{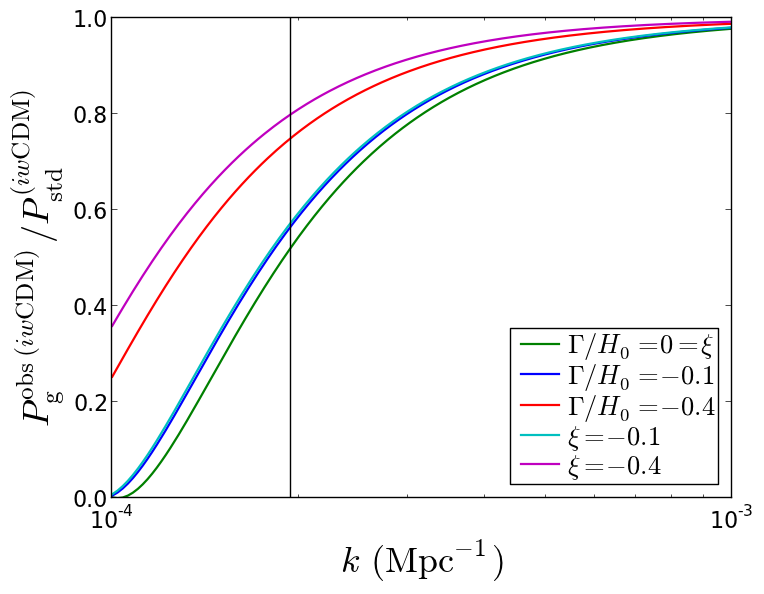}
\end{tabular}	
\caption{Ratios of the observed galaxy power spectrum $P^{\rm obs}_{\rm g}$ to the standard power spectrum $P^{\rm std}_{\rm g}$ along the line of sight ($\mu=1$), at $z=1$: for $w_x=-0.8$ ({\it left}) and $w_x=-1.1$ ({\it right}). }\label{fig:5}
\end{figure*}

\begin{figure*}
\begin{tabular}{cc}
\includegraphics[scale=0.45]{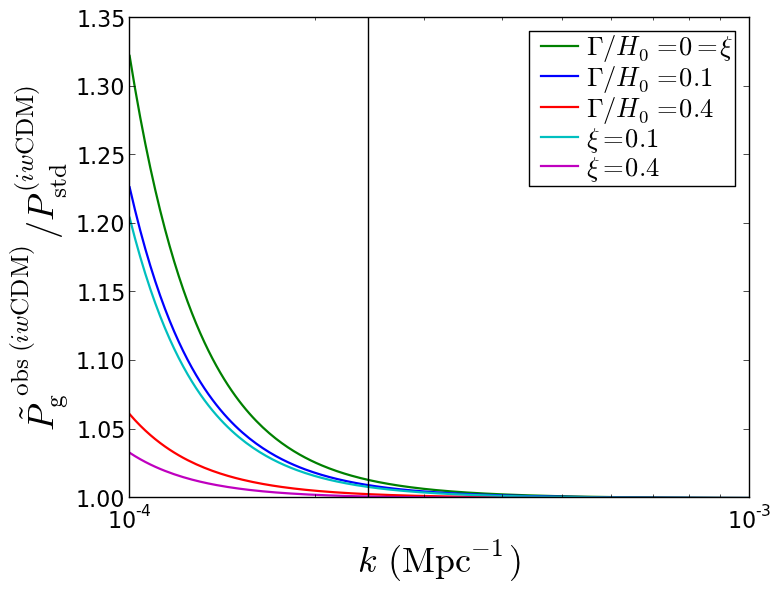} & 
\includegraphics[scale=0.45]{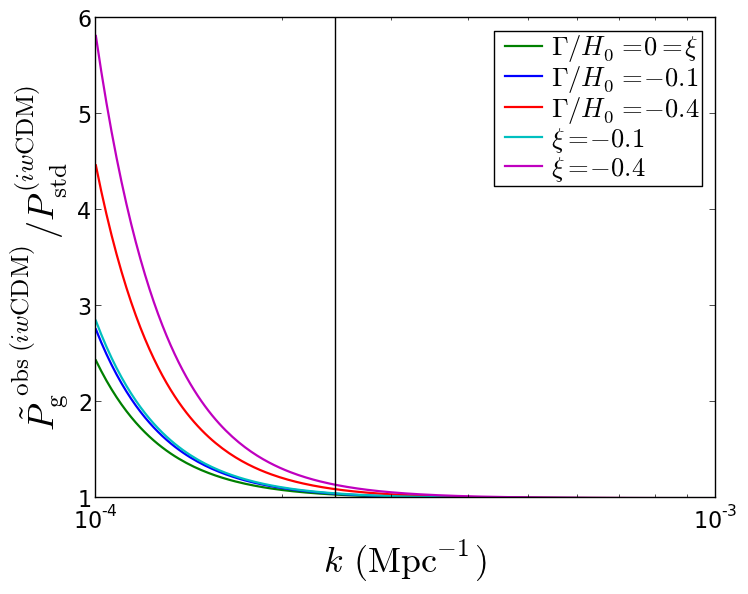} \\
\includegraphics[scale=0.45]{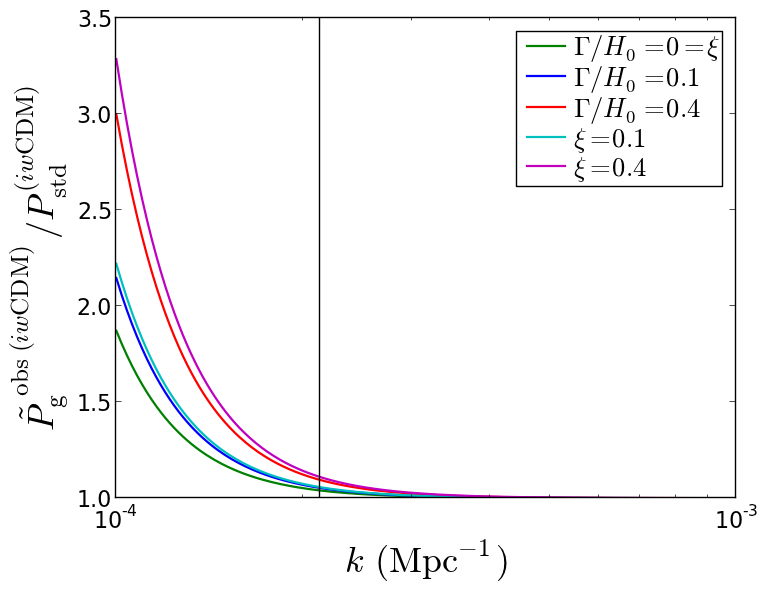} & 
\includegraphics[scale=0.45]{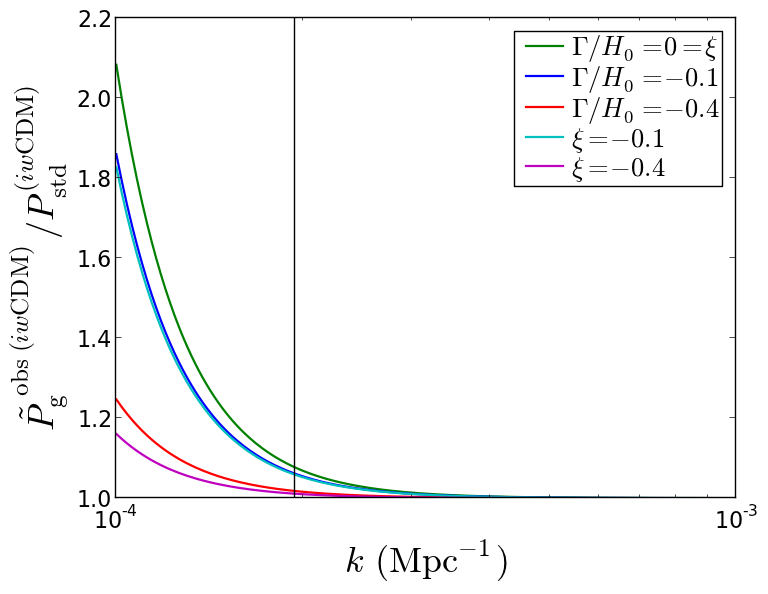} \\
\end{tabular}	
\caption{Ratios of $\tilde{P}^{\rm obs}_{\rm g}$, which is  $P^{\rm obs}_{\rm g}$ with the correlation between $\Delta^{\rm std}_{\rm g}$ and $\Delta^{\rm GR}_{\rm g}$ subtracted, to the standard power spectrum $P^{\rm std}_{\rm g}$, along the line of sight ($\mu=1$). Top panels have $z=0$, for $w_x=-0.8$ (top left) and $w_x=-1.1$ (top right). Similarly for $z=1$ in the bottom panels. }\label{fig:6}
\end{figure*}

From  \eqref{Delta:Obs}--\eqref{calB}, we obtain the power spectrum $P^{\rm obs}_{\rm g}$ of the observed galaxy overdensity in the conformal Newtonian gauge (we give only the real part):
\bea\label{Pk:obs}
{P^{\rm obs}_{\rm g} \over P_m}& =& \left(b + f\mu^2\right)^2 + 2\left(b + f\mu^2\right) {\cal A} {{\cal H}^2\over k^2}\nonumber\\ &&~ + {\cal A}^2 {{\cal H}^4\over k^4} -\mu^2 {{\cal B}^2}{{\cal H}^2\over k^2}. \label{Pk:Obs}
\eea
The standard power spectrum (\ie~the Kaiser approximation) is given by
\beq\label{Pk:std}
P^{\rm std}_{\rm g}(k,\mu,a) = \left[b(a) + f(k,a)\mu^2\right]^2 P_m(k,a).
\eeq
The matter power spectrum is obtained from \eqref{eq:lps:1} and \eqref{eq:lps:3}--\eqref{eq:lps:4} as \cite{Duniya:2013eta}
\beq\label{PmofK}
P_m(k,a) = \dfrac{9A^2}{50{\pi}^3H^n_0}\, k^n\, T(k)^2 \left[\dfrac{D_m(k,a)}{D_{\Phi0 }(k)}\right]^2.
\eeq
We note that the terms in the last line of \eqref{Pk:obs} correspond to the auto-correlations of $\Delta^{\rm GR}_{\rm g}$, and the last term in the first line corresponds to the cross-correlation between $\Delta^{\rm std}_{\rm g}$ and $\Delta^{\rm GR}_{\rm g}$. 

To compute the growth functions, we employ adiabatic initial conditions (see Appendix \ref{AICs}). For the goal of this work, it is reasonable to assume a constant comoving galaxy number density so that $b_e =0$. We also set (henceforth) the galaxy bias to $b=1$ for simplicity. We considered the case with the magnification bias 
\beq\label{mb1}
{\cal Q} = 1,
\eeq 
which corresponds to intensity mapping of the neutral hydrogen (HI) 21cm emission line \cite{Hall:2012wd,Duniya:2013eta,Peterson:2009ka}. In this case, the lensing convergence and time delay terms drop out of $\Delta^{\rm obs}_{\rm g}$, and the remaining integrated term, the ISW term, typically makes a negligible contribution. This gives a strong justification for our neglect of the integrated terms in  \eqref{Delta:GR}--\eqref{calB}.

Equation~\eqref{PmofK} shows that the induced changes in $P_m$ arising from the interaction will be imprinted via the ratio $D_m/D_{\Phi 0}$. This ratio is exactly unity on all scales in \LCDM\, while for dynamical DE it tends to 1 (by normalization) only on small scales $k \gg {\cal H}$, where the DE perturbations are negligible. This is illustrated in Fig.~\ref{fig:3} using $iw$CDM with $w_x=-0.8$ as an example. In this case, DM loses energy to DE, causing suppression of the matter growth function, which shows up in the matter power spectrum. Note that in the absence of interaction (as in the standard cosmologies), the clustering of DE  causes large-scale {\em suppression} in the matter power~\cite{Duniya:2013eta}. Moreover, the growth functions show that {\it Model~1} \eqref{Mod1:Q} is relatively more sensitive to the values of the interaction parameter, compared to {\it Model~2} \eqref{Mod2:Q}.

For $w_x=-1.1$ in $iw$CDM, the corresponding plots for $D_m/D_{\Phi 0}$ simply have all the curves reflected across that of \LCDM. 

In Fig.~\ref{fig:4} we present the line-of-sight GR corrected galaxy power spectrum $P^{\rm obs}_{\rm g}$ and the standard power spectrum $P^{\rm std}_{\rm g}$ in the Kaiser approximation, for the $iw$CDM models at $z=0$. 
On sub-Hubble scales,  \eqref{Pk:Obs} shows that
\beq\label{PofKApprox}
{P^{\rm obs}_{\rm g} \over P_m}\to \left(1 + f\right)^2~~\mbox{for}~k\gg{\cal H},
\eeq
\ie~$P^{\rm obs}_{\rm g} \to P^{\rm std}_{\rm g}$.

We also see that $iw$CDM gives a large-scale boost in galaxy power for both $\Gamma,~\xi>0$ ($w_x >-1$) and $\Gamma,~\xi<0$ ($w_x <-1$). As remarked above, there is less sensitivity to the values of $|\xi|$, so that {\it Model~2}~\eqref{Mod2:Q} predicts relatively larger amplitudes on super-horizon scales than does {\it Model~1}~\eqref{Mod1:Q}.

The large-scale boost in galaxy power in Fig.~\ref{fig:4} for $w$CDM arises purely from GR effects. The smaller boost (relative suppression) in power for $iw$CDM with $\Gamma,~\xi > 0$ (top left) comes from the fact that DM loses energy to DE, so that the higher the DM rate of loss of energy (\ie~larger $|\Gamma|,~|\xi|$), the more the power suppression. For $\Gamma,~\xi < 0$ (bottom left), where DE loses energy to DM, larger values of $|\Gamma|$ and $|\xi|$ give more boost relative to $w$CDM. 

Here, GR corrections in the galaxy power spectrum result in large-scale galaxy power enhancement. This implies that if we ignore the GR effects, \ie~if we do not subtract them in order to isolate the IDE effects, then we arrive at an incorrect estimate of the imprint of IDE on very large scales.

In \fig~\ref{fig:4} (right panels), we also show the corresponding ratios of the total galaxy power spectrum relative to the Kaiser approximation, at $z=0$. The ratios show that, for $\Gamma,~\xi > 0$ (top right), GR effects are suppressed relative to $w$CDM by the interaction. This is consistent with previous explanations above. Conversely, for $\Gamma,~\xi < 0$ (bottom right), GR effects are  enhanced relative to $w$CDM.

However, note that the total galaxy power spectrum contains not only the individual contributions of the standard Kaiser redshift-space distortion and GR terms, but also their cross-correlation. This cross-correlation makes a positive contribution at low $z$, and a negative contribution at high $z$ -- for the given magnification bias \eqref{mb1}. Hence, the ratios shown (at $z=0$) contain positive contributions of this cross correlation term, as well as the auto-correlation of the GR corrections. The ratios at $z=0$ are thus enhanced on horizon scales.

The GR effects in our IDE models, where we use Gaussian primordial perturbations, show a similar behaviour to the effects of primordial non-Gaussianity in non-interacting DE models (with $f_{\rm NL}>0$). The degeneracy between GR effects and primordial non-Gaussianity in the \LCDM\ model has been investigated by~\cite{Bruni:2011ta,Jeong:2011as,Maartens:2012rh,Yoo:2012se, Camera:2014bwa,Camera:2014sba}. It was recently shown by \cite{Hashim:2014rda} that IDE effects (in the case where GR effects are neglected) can be degenerate with primordial non-Gaussianity. 

In \fig~\ref{fig:5}, we show the ratios of the observed galaxy power spectrum to the Kaiser approximation at higher redshift, $z=1$.  For higher values of the interaction parameters, the case with $\Gamma,~\xi > 0$ (left) shows lower large-scale GR effects in comparison to the case with $\Gamma,~\xi < 0$ (right). For the magnification bias \eqref{mb1} (corresponding to HI intensity mapping), the observed line-of-sight power spectrum falls to zero for both non-interacting and interacting DE.

At higher $z$ the IDE effects are weaker, since the effects of DE in general are weaker at earlier times. By contrast, the GR effects are typically stronger at higher $z$ -- but with $\Delta^{\rm GR}_{\rm g}$ having negative amplitude. Hence its correlation with the Kaiser  term gives negative contribution in the power spectrum, thereby gradually reducing galaxy power on horizon scales. For completeness, we illustrate this phenomenon in Fig.~\ref{fig:6}, which shows the ratio of $\tilde{P}^{\rm obs}_{\rm g}$, which is  $P^{\rm obs}_{\rm g}$ with the correlation between $\Delta^{\rm std}_{\rm g}$ and $\Delta^{\rm GR}_{\rm g}$ subtracted, to $P^{\rm std}_{\rm g}$. In the top panels,  $z=0$, and $z=1$ in the bottom panels. On the left, $w_x>-1$ and $\Gamma,~\xi > 0$, and on the right, $w_x<-1$ and $\Gamma,~\xi < 0$. By comparing the top left  and top right panels, with the top right  and the bottom right panels of Fig.~\ref{fig:4}, respectively, we see that the correlation  between the GR and the Kaiser terms is positive  at low $z$. Similarly, by comparing the bottom left and the bottom right panels, with the left and right panels of Fig.~\ref{fig:5}, respectively, we see that the cross correlation is negative and of larger amplitude at $z=1$.

\section{Conclusion}\label{sec:concl} %

We investigated GR effects in the observed galaxy power spectrum in two IDE models, comparing with the corresponding standard non-interacting DE scenarios. We focused on the case of magnification bias ${\cal Q}=1$, corresponding to HI intensity mapping, and normalized the IDE power spectra to those of their non-interacting DE counterparts on small scales at today, \ie\ by requiring that they have the same  $\Omega_{m0}$ and  $H_0$. This isolates the deviations arising from GR effects and IDE on very large scales.

We find that if the GR effects are disregarded, \ie~if they are not subtracted in order to isolate the IDE effects, then we arrive at an incorrect estimate of the imprint of IDE on horizon scales. This could lead to a bias in constraints on IDE on the given scales.

We also found that at low $z$, the correlation between the GR term and the (standard) Kaiser redshift-space distortion term has a positive contribution in the galaxy power spectrum, while at high $z$, this term gives a negative contribution that grows with increasing $z$.

Future wide and deep-field surveys may be able to disentangle any possible IDE effects from GR effects by comparing the observed power at low and high $z$. Detecting super-Hubble effects will be challenging because of cosmic variance. However, if the multi-tracer method~\cite{Seljak:2008xr} can be applied, cosmic variance can be reduced enough for detection of  these effects \cite{Yoo:2012se}.

\[\]{\bf Acknowledgements:} %
We thank Kazuya Koyama for useful comments. This work was supported by the South African Square Kilometre Array Project and the South African National Research Foundation. DB was also supported by the Deutsche Forschungsgemeinschaft through the Transregio 33, The Dark Universe. RM was also supported by the UK Science \& Technology Facilities Council (grant no. ST/K0090X/1).

\appendix 

\section{Adiabatic initial conditions}\label{AICs}%
We use the Einstein de Sitter initial condition $\Phi'(a_d)=0$, given that $\Omega_x(a_d)\ll 1$. Adiabatic initial conditions are imposed by the vanishing of the relative entropy perturbation $S_{xm}$, 
\beq \label{adic}
S_{xm}(a_d) = 0, \quad S_{xm} \equiv {\delta_x \over 1 +w_{x,\rm eff}} - {\delta_m \over 1 +w_{m,\rm eff}},
\eeq
and by the equality of velocities, 
\beq
V_x(a_d) = V_m(a_d).
\eeq
Then $(1 +w_{m,\rm eff}(a_d))\Delta_x(a_d) = (1 +w_{x,\rm eff}(a_d))\Delta_m(a_d)$. Together with \eqref{Poisson:Eq} and \eqref{Phi:Eqn}, this leads to the initial DM and DE fluctuations given by
\bea\label{Delta_mi}
\Delta_m(k) &=& \dfrac{-2k^2\left(1+w_{m,\rm eff}\right)}{3{\cal H}^2\left(1 + \Omega_m w_{m,\rm eff} + \Omega_x w_{x,\rm eff}\right)}  \Phi_d(k), \quad\\ \label{Delta_xi}
\Delta_x(k) &=& \dfrac{-2k^2\left(1+w_{x,\rm eff}\right)}{3{\cal H}^2\left(1 + \Omega_m w_{m,\rm eff} + \Omega_x w_{x,\rm eff}\right)} \Phi_d(k),\quad\\ \label{V_mi:V_xi}
V_x(k) &=& V_m(k) = \dfrac{-2}{3{\cal H} \left(1 + \Omega_x w_x\right)} \Phi_d(k),
\eea
where $w_{m,\rm eff}$ and $w_{x,\rm eff}$ are given by  \eqref{eff:EoS} and $\Phi_d(k)$ is given by  \eqref{phid}.

%

\end{document}